\documentclass[aps,prb,showpacs,floatfix,amsmath,amssymb,
               preprint]{revtex4-1}
\usepackage{color}
\usepackage{graphicx}
\usepackage{natbib}
\usepackage{epsfig}
\usepackage{setspace}
\usepackage{amsmath}
\usepackage{amssymb}
\usepackage{verbatim}
\usepackage{bibentry}

\begin{document}
\title{Renormalization group study of quantum criticality in gapped Kondo model}
\author{Rukhsan Ul Haq  and Anirban Sharma}
\affiliation{Theoretical Sciences Unit \\ Jawaharlal Nehru Center for Advanced \\Scientific Research Bangalore, India}
\begin{abstract}
Quantum impurity models are the prototypical examples of quantum many-body dynamics which manifests in  their spectral and transport properties.  Single channel Anderson(and Kondo model) leads to the Fermi liquid ground state in the strong coupling regime which corresponds to a stable infrared fixed point at which the quantum impurity gets completely screened by the conduction electrons. Quantum impurity models with non-trivial density of states exhibit quantum phase transition and this quantum criticality lies in the universality class of local quantum critical systems. In this paper, we report first study of the flow equation renormalization of gapped Kondo model which has  gapped density of states, the gap being at the Fermi level.  Flow equation renormalization group method  has proved to be one of the very robust renormalization methods to study Kondo physics both in equilibrium as well as in non-equilibrium. Here we have shown that this method can also be employed to study local quantum criticality. We have calculated the flow equations for the Kondo coupling and solved them for various values of the gap parameter and we find that there is suppression of Kondo divergence as gap is increased which signifies that as gap is increased, renormalization flow goes away from the strong coupling fixed point. We have also calculated the spin susceptibility and we find that as gap is increased, susceptibility goes over to the Curie behaviour and hence confirming the renormalization flow towards the local moment fixed point.
\end{abstract}
\maketitle
\section{Introduction}
Quantum impurity models have played a crucial role in the understanding of the strongly correlated electron systems. They have also led to the development of many non-perturbative methods which especially include numerical renormalization group method which was developed by Wilson for solving the ``Kondo problem" and was later on applied to study the Anderson impurity model as well. Kondo model is the simplest model which leads to the non-perturbative Kondo physics in which quantum many-body Kondo singlet is formed due to the antiferromagnetic interaction between magnetic impurity and conduction electrons. Kondo singlet formation takes place below Kondo temperature which is the universal scale which distinguishes between strong coupling and weak coupling regimes of the model. One of the important features of Kondo physics is that Kondo scale gets dynamically generated due to the strong correlation induced renormalization of the bare parameters. In the strong coupling regime of the Kondo model, there is quenching of the magnetic moment and hence the ground state of the model is local Fermi liquid. However, multichannel generalization of Kondo model leads to the non-Fermi liquid ground states. Recently there has been lot of interest in quantum criticality in quantum impurity models and this quantum criticality has been called local quantum criticality. Local quantum criticality can not be understood within the Landau-Wilson paradigm of the phase transitions. One of the important aspects of local quantum criticality is that it can be realized in quantum dots. 

Though the Kondo model leads to Fermi liquid ground state it was found in \cite{Fradkin} that Kondo model with a pseudo-gap in the density of states can
 exhibit quantum phase transition with gap being the tuning parameter. Pseudo-gap model has been extensively studied based on numerical renormalization group methods as well as other perturbative renormalization group methods. In this paper we study gapped Kondo model which has a gap in the density of states at the Fermi level. This gapped Kondo model can be obtained as an effective Hamiltonian from gapped Anderson impurity model using Schrieffer-Wolff transformation. This model has been studied before using poor man scaling method, $\frac{1}{N}$ expansion method, numerical renormalization group method and more recently local moment method has also been applied to study this model\cite{Saso,Sakai,Ogura,Chen,Moca,Pinto,Galpin,Logan}. We have applied flow equation renormalization method to study this model. Flow equation method allows us not only to calculate the flow equations for the Kondo coupling but also gives access to the dynamic quantities like dynamic spin susceptibility for Kondo model. For the gapped Kondo model,we have not only calculated the flow equations which have been solved numerically, but we have also calculated the dynamic quantities like spin susceptibility. The rest of the paper is organized as follows: In next section we have given the description of the model and the renormalization method which we have employed. In  section \textbf{3} we calculate the flow equations and solve them numerically for various values of the gap parameter. In the next section \textbf{4} we have calculated the renormalization flow of the spin operator which gives us access to the dynamical spin susceptibility from which other static quantities can be readily calculated. Finally we summarize and discuss our results in the conclusion section.
\section{Model and Methods}
%\subsection{Flow Equation Method treatment of Kondo Model}
In this section we will carry out flow equation renormalization of gapped Kondo model. We write gapped Kondo model as:
\begin{equation}
H(l)=H_{0}+H_{int}(l)
\end{equation}
where
\begin{equation}
H_{0}=\sum_{t,\alpha}\epsilon_{t}c^{\dag}_{t\alpha}c_{t\alpha}
\end{equation}
\begin{equation}
H_{int}(l)=\sum_{t't}J_{t't}(l):S.s_{t't}:
\end{equation}
Here t and t' are general indices which in this case represent momenta. One should note that normal ordering prescription which is used for quantum many-body systems has been incorporated. The conduction electron spin desnity is given by:
\begin{equation}
s_{t't}=\sum_{\alpha,\beta}c^{\dag}_{t'\alpha}\frac{\sigma_{\alpha\beta}}{2}c_{t\beta}
\end{equation}
%\subsubsection{1-loop calculation}
%Density of states for the conduction band electrons has the following form:
%\begin{align}
%\rho(\epsilon_{k})=&{0   -\delta < \epsilon_{k} < \delta \nonumber \\
%                   & \frac{1}{2D}}
%\end{align}
                              
First we calculate the generator for flow equations of Kondo model.
\begin{align}
&\eta^{1}(l)=[H_{0},H_{int}(l)] \\
&\eta^{1}(l)=\sum_{t't}\eta_{t't}^{1}(:S.s_{t't}:) \label{eta-1}
\end{align}
where $\eta_{t't}^{1}(l)=(\epsilon_{t'}-\epsilon_{t})J_{t't}(l)$.
The commutator of generator with the diagonal part of the Hamiltonian is
\begin{equation}
[\eta^{1}(l),H_{0}]= -\sum_{t't}(\epsilon_{t'}-\epsilon_{t})^{2}J_{t't}(l):S.s_{t't}:
\end{equation}
Evaluating the commutator of the generator with the Kondo interaction taking normal ordering of the fermionic operators into consideration, we obtain the flow equations for the gapped Kondo model.
The flow equations for Kondo coupling till first loop is:
\begin{align}
&\frac{d J_{t't}}{dl}=-(\epsilon_{t'}-\epsilon_{t})^{2}J_{t't}+\nonumber\\
&\sum_{v}(\epsilon_{t'}+\epsilon_{t}-2\epsilon_{v})J_{t'v}J_{vt}(n(v)-1/2)
\end{align}
Unlike poor man scaling where Kondo coupling is momentum independent and hence there is only one scaling equation for isotropic Kondo model, the flow equation given above is actually a system of equations corresponding to different momenta. The set of equations in non-linear and coupled. Hence analytical solution is not possible except in some special cases.
\subsection{Numerical Solution of Flow equations of Gapped Kondo Model}
Flow equations are coupled non-linear differential equations. They generally can not be solved analytically except in some special limits e.g. in the infrared limit when the momentum of the coupling constants is restricted to be close to Fermi level. In this case, flow equations  reproduce the results of the conventional scaling methods like poor man scaling. Flow equations can be solved by numerical methods like as Runga-Kutta methods. The number of flow equations to be solved scales as $O(N^{2})$ where N is the number of energy states of the conduction band in case of Kondo model. Since flow equations are renormalization flows and hence they meet many different energy scales during the unitary flow and hence become stiff also. In this section, we solve the flow equations of Kondo model numerically. We have used DOPRI5 Fortran subroutine as the solver. DOPRI5 is based on fifth order Runga-Kutta method.
\begin{figure}[h]
\centering
\includegraphics[scale=0.6]{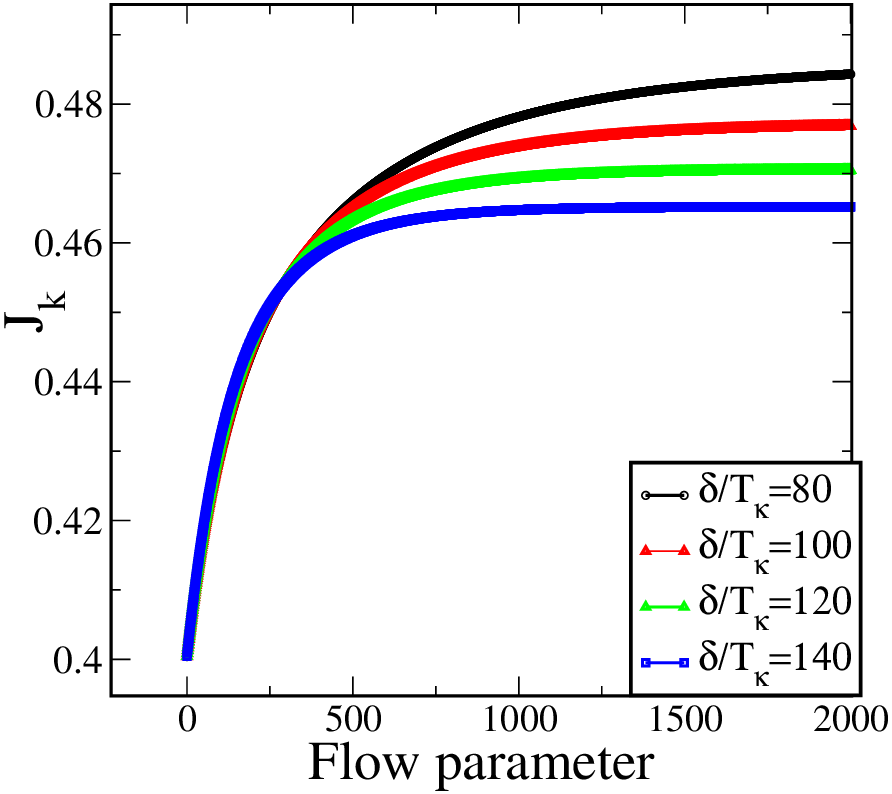}
\caption{Kondo coupling has been plotted versus flow parameter for increasing values of gap.}\label{Kondo_flow}
\end{figure} 
In the infrared limit, flow equations recover poor man scaling equation of Kondo model. Kondo coupling grows logarithmically and finally diverges at Kondo scale. Numerical solution of flow equations gives access to all the Kondo couplings for different momenta. However to see the divergence of the Kondo coupling we did infrared parametrization numerically by restricting momenta to Fermi level($\epsilon_{k}=0=\epsilon_{k'}$).  In the isotropic Kondo model we find that Kondo coupling diverges and we have found that in the numerical solution of the flow equations for the isotropic Kondo model. However, in presence of the gap, Kondo divergence gets cut-off as has been shown in  Figure~\ref{Kondo_flow}. As the gap in increased, the flow of Kondo coupling gets flatter and saturates at some finite value, which signifies that the fixed point is not Kondo fixed point.
\section{Flow of the spin operator}
In this section we will calculate the flow equation for Kondo impurity spin operator and then calculate the dynamic spin susceptibility from the numerical solution of the  flow equations.
\begin{equation}
\frac{dS^{a}(l)}{dl}=[\eta_{l},S^{a}(l)]
\end{equation}
We will use the one loop generator, given in equation~\ref{eta-1} to evaluate the above commutator. We use the  following ansatz for spin operator:
\begin{equation}
S^{a}(l)=h(l)S^{a}+i\sum_{u'u}\gamma_{u'u}(l):(S\times s_{u'u})^{a}:
\end{equation}
The flow equations for the coefficients are
\begin{align}
&\frac{d h}{dl}=\sum_{t't}(\epsilon_{t'}-\epsilon_{t})J_{t't}\gamma_{tt'}n(t')(1-n(t)) \\
&\frac{d\gamma_{t't}}{dl}=h(\epsilon_{t'}-\epsilon_{t})J_{t't}\nonumber\\
&-\frac{1}{4}\sum_{u}((\epsilon_{u'}-\epsilon_{u})J_{t'u}\gamma_{t'u})(1-2n(u)))
\end{align}
%\subsection{Spin correlation function and Dynamical susceptibility}
Using the above formalism we can calculate spin-spin correlation function.
\begin{align}
C(t)=\frac{1}{2}\langle{S^{z}(0),S^{z}(t)}\rangle 
\end{align}
Spin-spin correlation function as function of the frequency is given as:
\begin{align}
C(\omega)&=-\frac{\pi}{Z(\beta)}\sum_{n}e^{-\beta E_{n}}\sum_{tt'}\sum_{uu'}\gamma_{tt'}(l\rightarrow \infty)\gamma_{u'u}(l\rightarrow\infty)\nonumber \\
&\times \langle n\vert:(S\times s_{t't})^{z}::(S\times s_{u'u})^{z}:\vert n\rangle\nonumber\\
&\times (\delta(\omega -\epsilon_{u'}+\epsilon_{u})+\delta(\omega +\epsilon_{u'}-\epsilon_{u}))
\end{align}
\begin{align}
C(\omega)= \frac{\pi}{4}\sum_{u}\gamma^{2}_{\epsilon_{u}+\omega,\epsilon_{u}}(l\rightarrow \infty)\times (n(\epsilon_{u})(1-n(\epsilon_{u}+\omega)(1-n(\epsilon_{u})
\end{align}
The quantity which we are interested in and which we have calculated is imaginary part of the dynamic spin susceptibility $\chi(\omega)$. Fluctuation-dissipation theorm relates $\chi(\omega)$ to the spin-spin correlation function $C(\omega)$ calculated above.
\begin{align}
\chi(\omega)=tanh\left(\frac{\omega}{2T}\right)C(\omega)
\end{align}
\begin{figure}[h]
\centering
\includegraphics[scale=0.55]{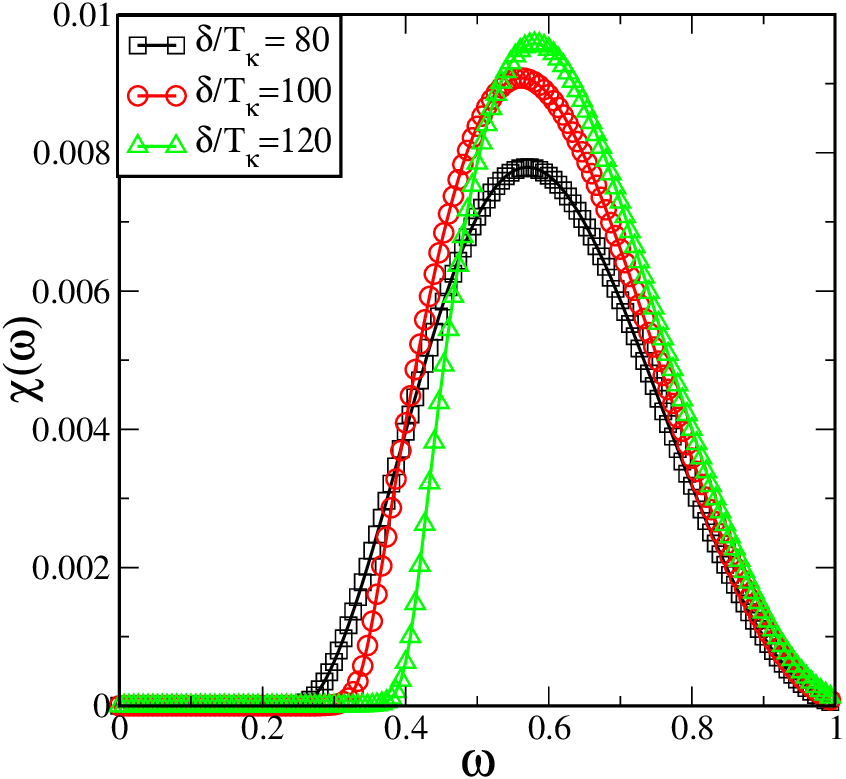}
\caption{ Spin susceptibility of Gapped Kondo model is plotted as function of frequency for various values of gap parameter.}\label{Kondo_chi_gap}
\end{figure}
Dynamic spin susceptibility is plotted in Figure~\ref{Kondo_chi_gap} for various values of the gap parameter. One important difference with respect to isotropic Kondo model is that spin susceptibility is until some value of the frequency which is related to the value of the gap. In contrast for the isotropic Kondo model, zero frequency spin susceptibility shows the linear behaviour as expected from Korringa relations. We also see in Figure~\ref{Kondo_chi_gap} that as gap is increased, spin susceptibility is increased.

\begin{figure}[h]
\centering
\includegraphics[scale=0.55]{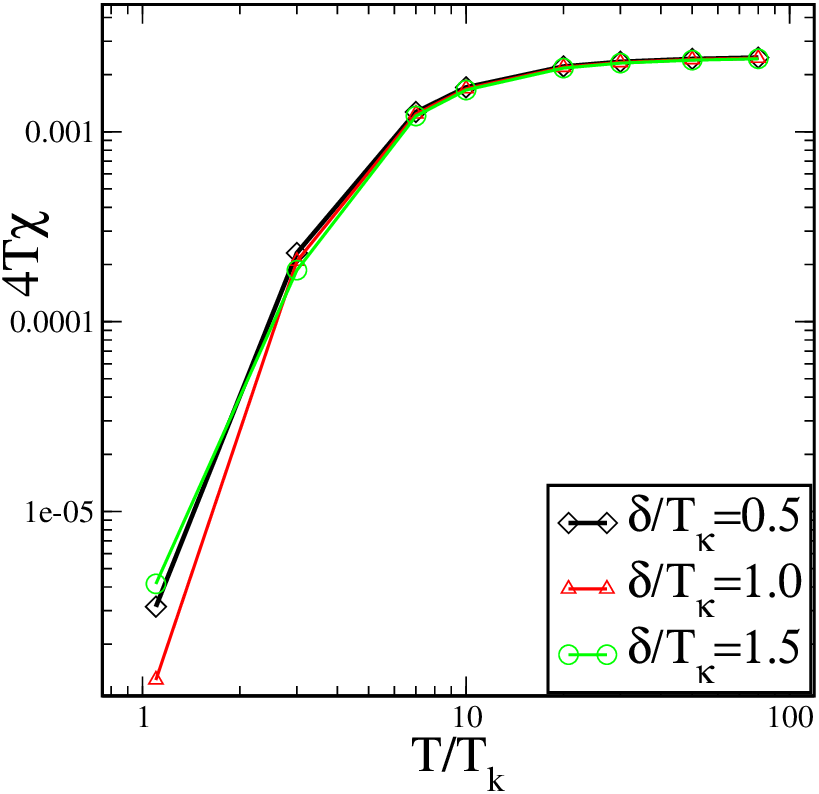}
\caption{ Static spin susceptibility of Gapped Kondo model is plotted as function of temperature for various values of gap.}\label{chi0_gap}
\end{figure}

One another important quantity for the Kondo model is the static spin susceptibility which is related to the inverse Kondo temperature. Static spin susceptibility can be readily calculated from the dynamic susceptibility by integration. We have calculated static spin susceptibility for gapped Kondo model as function of temperature for various values of the gap parameter as shown in Figure~\ref{chi0_gap}. Our results are in agreement with earlier results based on numerical renormalization group\cite{Chen} and quantum monte carlo calculations\cite{Saso}.

%As we decrease Kondo coupling and hence Kondo effect becomes weaker, spin susceptibility gets enhanced. We want to point out that we obtained these curves with out any broadening which has been done in\cite{Kehrein1,Kehrein2}.
%These curves are the universal curves for the spin susceptibility for the isotropic Kondo model\cite{Kehrein1}. 
\section{Conclusion}
In this paper we have applied flow equation renormalization method to gapped Kondo model to explore the quantum phase transition in this model. This model has a gap in the density of conduction electrons at the Fermi level. Gap is the tuning parameter which is used to go from metallic ground state at the strong coupling fixed point to the local moment ground state. We have calculated the flow equations for the model which have been solved numerically. We find that as we increase the gap, Kondo divergence gets cut off and hence the systems flows away from the strong coupling fixed point. To confirm that the flow is towards the local moment fixed point we have calculated the static susceptibility and we find that at high temperature susceptibility shows local moment behaviour. We have also calculated dynamic spin susceptibility and we find that as the gap is increased, spin susceptibility gets enhanced. Our results are in agreement with earlier calculations based on numerical renormalization group method and quantum monte carlo simulations. Ours is the first study where flow equation renormalization group method has been used to study the quantum phase transition in gapped Kondo model. We not only have explored the new aspects of the quantum  criticality by calculating the dynamic quantities but our study also shows that flow equation method can be employed to study the quantum criticality in impurity models.

\end{document}